\documentclass[fleqn,12pt,twoside]{article}
\usepackage{espcrc1}

\usepackage{graphicx}
\usepackage{epsfig}
\usepackage[figuresright]{rotating}

\newcommand{\AmS}{{\protect\the\textfont2
  A\kern-.1667em\lower.5ex\hbox{M}\kern-.125emS}}

\title{Probing the QCD Equation of State\thanks{Work supported in part by BMBF, GSI and by the European Commission under contract HPMT-CT-2001-00370.}}

\author{R. A. Schneider\address[TUM]{Physik-Department, Technische Universit\"at M\"unchen, 85747 Garching, Germany}\address[ECT]{ECT*, Villa Tambosi, 38050 Villazzano (Trento), Italy},
        T. Renk\addressmark[TUM]\addressmark[ECT],
	M. Thaler\addressmark[TUM],
	A. Polleri\addressmark[TUM]\addressmark[ECT] and 
	W. Weise\addressmark[TUM]\addressmark[ECT]}

\begin{document}
%
\maketitle

\begin{abstract}
We propose a novel quasiparticle interpretation of the equation of state of deconfined QCD at finite temperature. Using appropriate thermal masses, we introduce a phenomenological parametrisation of the onset of confinement in the vicinity of the phase transition. Lattice results of bulk thermodynamic quantities are well reproduced, the extension to small quark chemical potential is also successful. We then apply the model to dilepton production and charm suppression in ultrarelativistic heavy-ion collisions.
\end{abstract}
%
%
%

\section{Introduction}
%
QCD is expected to undergo a transition from a confined hadronic phase to a deconfined partonic phase, the quark-gluon plasma (QGP), at a temperature of $T_c \sim 170$ MeV \cite{LAT1}. A central quantity of matter in thermal equilibrium is the Helmholtz free energy from
which the pressure $p$, energy density $\epsilon$ and entropy density $s$ -- which are important
ingredients for the description of ultrarelativistic heavy-ion collisions (URHIC) -- can be derived. A
first-principles understanding of the equation of state (EOS) of hot QCD is therefore of great
interest, also in order to reliably identify and calculate experimental signatures of that elusive
state. Perturbative results on the EOS are available up to order $\mathcal{O}(g_s^5)$ \cite{ZK95}, but show bad convergence for all temperatures of interest. Non-perturbative methods such as lattice QCD calculations hence become
mandatory. From these numerical simulations the EOS of a pure gluon plasma is known to high accuracy
\cite{BE96}, and there are first estimates for the continuum EOS of systems including quarks \cite{FPL00,FK02}.

Various interpretations of the lattice data have been attempted, most prominently as the EOS of a gas
of quark and gluon quasiparticles. In a phenomenological framework, quarks and gluons are simply
treated as non-interacting, massive quasiparticles \cite{PKS00}. More recently, a quasiparticle
description of QCD thermodynamics has been derived in a more rigorous treatment using resummed hard thermal loop (HTL)
perturbation theory \cite{BI01}. Employing the full HTL spectral representions, the resulting EOS can
be matched to lattice data down to temperatures $T \sim 3 \ T_c$; below that temperature,
non-perturbative physics not amenable in an expansion in $g_s$ becomes important. Unfortunately, as
evident from figure \ref{figure1} (left panel), current experiments only probe that very temperature
regime where the underlying physics, the confinement and chiral symmetry breaking mechanism, is not
sufficiently well understood. Phenomenological models incorporating as much physics as is known are
therefore necessary. Here, we propose a new quasiparticle model of the QGP that incorporates a parametrisation
of confinement close to $T_c$, supplemented by thermal masses compatible with lattice results. Details
can be found in \cite{RAS}.
%
\section{Quasiparticles and confinement}
%
Consider a SU(3) gluon plasma. From asymptotic freedom, we expect that at very high temperatures the plasma
consists of quasifree gluons. As long as the spectral function of the thermal
excitations at lower temperatures resembles qualitatively this asymptotic form, a
gluonic quasiparticle description is expected to be applicable. The dispersion equation of the gluonic quasiparticles reads
\begin{equation}
\omega^2_k \simeq k^2 + m^2_*(T), \label{disp_rel}
\end{equation}
where $m_*(T)$ acts as an effective mass generated dynamically by the interaction of the gluons with
the heat bath background.
%
\begin{figure}[bth]
%
\begin{center}
\vspace{-0.7cm}
\epsfig{file=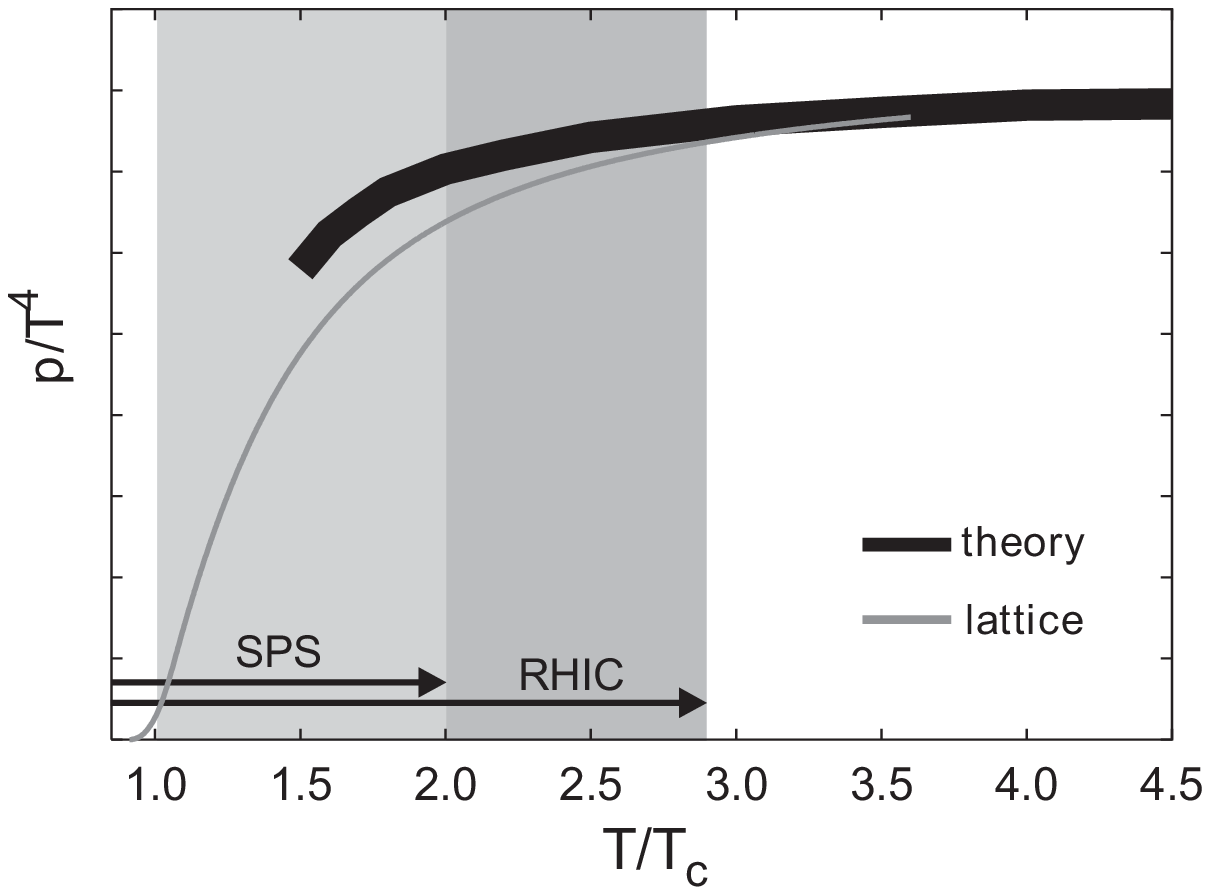,width=7.5cm} \epsfig{file=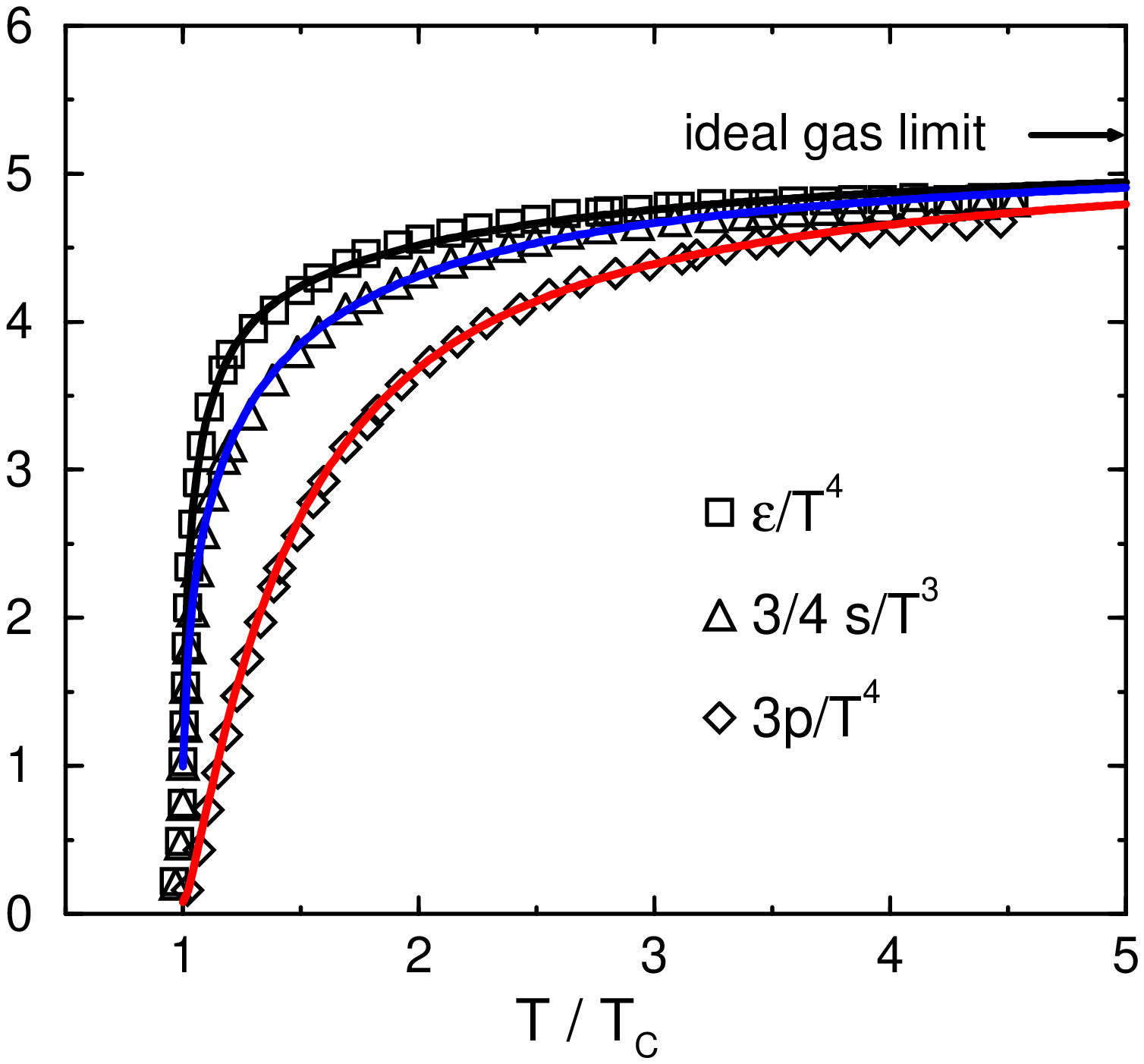,width=7.25cm}
\vspace{-0.7cm}
\caption{Left panel: Sketch of theory vs. lattice and the temperatures probed by current experiments.
Right panel: Normalised $\epsilon$, $s$ and $p$ (solid lines) in a quasiparticle approach \cite{RAS}, compared to continuum extrapolated SU(3)
lattice data (symbols) \cite{BE96}.} \label{figure1}
\end{center}
\vspace{-0.8cm}
\end{figure}

However, the picture of a simple massive gas is presumably not appropriate close to $T_c$ because the
driving force of the transition, the confinement process, is not taken into account. This physics has
to be incorporated phenomenologically. Below $T_c$, the relevant degrees of freedom in pure SU(3) gauge theory are heavy,
colour singlet glueballs. Approaching $T_c$, deconfinement sets in and the gluons are liberated,
followed by a sudden increase in entropy and energy density. Conversely, when approaching the phase
transition from above, the number of thermally active degrees of freedom is reduced due to the onset of
confinement. As $T$ comes closer to $T_c$, an increasing number of gluons gets trapped in glueballs
which {\em disappear} from the thermal spectrum: since $m_{GB} \sim 1.5 $ GeV and $T_c \sim 270$ MeV, glueballs are simply too heavy to become thermally excited in the temperature
range under consideration (up to $5 \ T_c$). So, all confinement does statistically on a {\em large} scale is to cut down the number of thermally active gluons as the temperature is lowered. This effect can be included in the quasiparticle picture by modifying the distribution function of the gluons by a temperature-dependent {\em confinement factor $C(T)$}: $f_B(\omega_k)
\rightarrow C(T) f_B(\omega_k)$.

%
\begin{figure}[tbh]
%
\begin{center}
\vspace{-0.4cm}
\epsfig{file=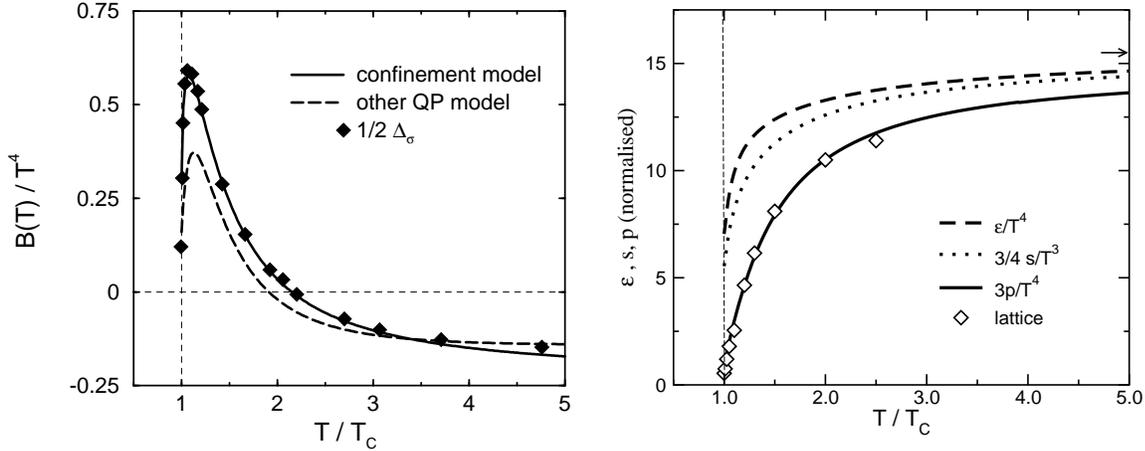,width=8.2cm}\raisebox{4mm}{\epsfig{file=2_1_flav.eps,width=7.0cm}}
\vspace{-0.7cm}
\caption{Left panel: The function $B(T)$. Symbols show $\frac{1}{2}\Delta_\sigma$ \cite{BE96}, the
dashed line displays $B(T)$ in other quasiparticle models \cite{PKS00}. Right panel:  Normalised
$\epsilon$, $s$ and $p$ (solid lines) for $N_f = 2 + 1$. The data points are from the lattice simulation in \cite{FK02}.} \label{figure2}
\end{center}
\vspace{-0.8cm}
\end{figure}
%
%

To become quantitative, we have to specify the thermal masses $m_*(T)$ entering (\ref{disp_rel}). Based
on the observation that the Debye screening mass $m_D$ evaluated on the lattice shows approximate
critical behaviour \cite{KK00}, in accordance with a weakly first order phase transition and in
contrast to perturbative results, we parametrise
\begin{equation}
m_*(T) \sim G_0  T \left(1 -  \frac{T_c}{T} \right)^{\beta}. \label{m_g}
\end{equation}
Consider now the entropy of a gas of massive gluons with such a dropping $m_*(T)$. The result for
$s(T)$ will clearly overshoot the lattice entropy because light masses near $T_c$ lead to an increase
in $s(T)$. However, since the entropy is a measure for the number of active degrees of freedom, the
difference may be accounted for by the aforementioned confinement process as it develops when the
temperature is lowered towards $T_c$. The explicit temperature dependence of the confinement factor
$C(T)$ can be obtained simply as the ratio of the lattice entropy and the entropy calculated with a
dropping input gluon mass, and $C(T)$ again shows near-critical behaviour: $C(T) \sim (1 - T_c/T)^\gamma$.
Thermodynamical quantities like the energy density can now be calculated as
\begin{equation}
\epsilon(T) = 16  \int \frac{d^3 k}{(2\pi)^3}  \left[ C(T) f_B(\omega_k) \right] \
\omega_k + B(T).
\end{equation}
The function $B(T)$ is not an independent quantity, but uniquely determined by $m_*(T)$, $C(T)$ and
their $T$-derivatives. It is necessary to maintain thermodynamical self-consistency and can be
interpreted as the thermal energy density of the vacuum. The explicit expressions for $B(T)$, the
pressure and the entropy density can be found in \cite{RAS}. In figure \ref{figure1} (right panel), we
compare $\epsilon$, $s$ and $p$ with lattice data, as a function of $T$. We achieve a good and economic
parametrisation. Figure \ref{figure2} (left panel) shows the
function $B(T)$ and the spacelike plaquette expectation value $\Delta_\sigma$, as measured on the
lattice, that can be related to the thermal chromomagnetic condensate $\langle \mathbf{B}^2 \rangle_T$. We find the simple relation
$$
B(T) = \frac{1}{2}\Delta_\sigma(T) T^4 = - \frac{11 \alpha_s}{8\pi} \langle \mathbf{B}^2 \rangle_T +
\frac{1}{4} \langle G^2 \rangle_{T=0}
$$
with the zero-temperature condensate $\langle G^2 \rangle_{T=0}$. This correlation between $B(T)$ and
$\langle \mathbf{B}^2 \rangle_T$ may hint at a deeper connection between $B(T)$ as a carrier of
non-perturbative effects, and the magnetic condensate. After all, $B(T)$ represents the thermal energy
of the (non-trivial) Yang-Mills vacuum. 
%
\section{Dynamical quarks}
%
The extension of the mechanism presented so far to systems with dynamical quarks is not
straightforward. Simulations of fermions on the lattice are still plagued by problems. However, when
plotting the lattice pressure, normalised to the ideal gas value, for the pure gauge system and for
systems with 2, 2+1 and 3 quark flavours, it is found that the QCD EOS shows a remarkable flavour {\em
independence} when plotted against $T/T_c$. The flavour dependence is then well approximated by a term
reminiscent of an ideal gas $p(T, N_f) \sim \left(16 + 10.5 \ N_f \right) \tilde{p}(T/T_c)$ with a
universal function $\tilde{p}(T/T_c)$ \cite{FK01}. This hints at a confinement mechanism being only
weakly flavour-dependent, and hence we assume that the function $C(T)$ acts in a universal way on
quarks and gluons. Figure \ref{figure2} (right panel) shows pressure, energy and entropy density for
two massless quark flavours and a heavy strange quark, compared to lattice data from a simulation with slightly different masses. The agreement especially close to $T_c$ is certainly encouraging.

\begin{figure}[bth]
%
\begin{center}
\vspace{-0.9cm}
\epsfig{file=quark_number_density.eps,width=7cm} \raisebox{59mm}{\epsfig{file=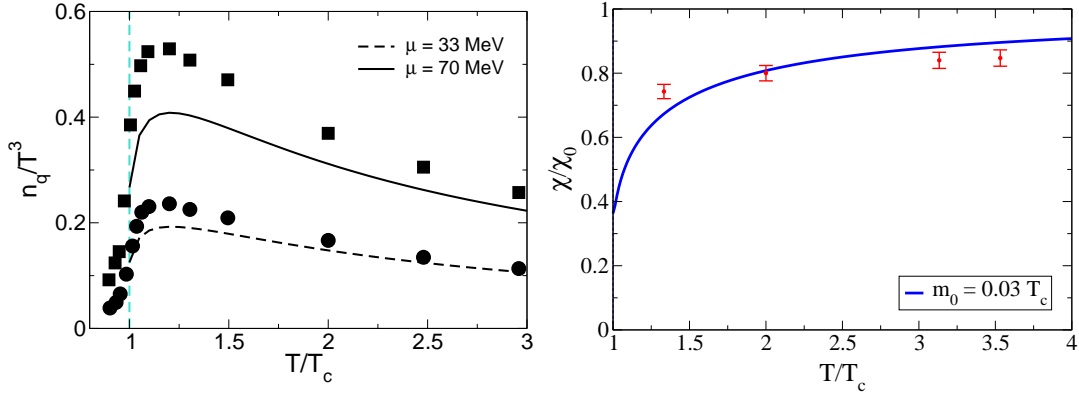,width=6.1cm, angle = -90}}
\vspace{-0.9cm}
\caption{Left panel: Net quark number density $n_q/T^3$ at small $\mu$ for 2+1 flavours. Data points are continuum estimates of lattice simulations with $m_{u,d} \simeq 65$ MeV and $m_s \simeq 135$ MeV \cite{FK02}. Right panel: Quark number susceptibility for 2+1 flavours, normalised to the ideal gas value. The data points are continuum estimates of the lattice simulation of  \cite{Gavai:2002}.} \label{figure3}
\end{center}
\vspace{-0.8cm}
\end{figure}

For small quark chemical potential $\mu$, it is reasonable to assume that $C(T)$ is only weakly $\mu$-dependent. The net quark density then takes the form
\begin{equation}
n_i(T, \mu) = 3 \int \frac{d^3 k}{(2\pi)^3} C(T) [ f^+_D - f^-_D] \quad \mbox{with} \quad f_D^\pm = \frac{1}{\exp([\omega_k \mp \mu_i]/T) + 1},
\end{equation}
where $i$ labels the flavour. Figure \ref{figure3} (left panel) shows $n_q/T^3$ as a function of $T/T_c$, compared to preliminary continuum-extrapolated lattice data \cite{FK02}. The general shape is well reproduced, which supports the validity of the quasiparticle assumption even close to $T_c$. Another quantity that tests the $\mu$-dependence is the quark number susceptibility $\chi(T, \mu)_{ij} = \partial n_i / \partial \mu_j$. In figure \ref{figure3} (right panel), we plot $\chi = \chi(T, \mu = 0)_{ii}$, normalised to the ideal gas value $\chi_0 = N_f T^2$, along with corresponding lattice data \cite{Gavai:2002}. Again, the agreement is satisfactory. 
%
%
%
%
\section{Dilepton rates}
%
%
We apply the model now to dilepton production in URHIC where it enters in two ways: first, the
dilepton emissivity of a static hot source,
\begin{equation}
\frac{d\mathcal{N}_{ee}}{d^4x d\omega d^3 k} \sim \frac{\rho_V(\omega, k; T)}{\exp(\omega/T) -1},
\label{rate}
\end{equation}
is proportional to the vector spectral function $\rho_V$ that is calculated from the photon self
energy. By construction, the quasiparticle $q\bar{q}$-loop is the only contribution in the QGP phase. In the hadronic phase, we use vector meson dominance to couple the photon to the $J^P = 1^-$ mesons $\rho$, $\omega$ and $\phi$, taking into account temperature and baryon density effects \cite{Renk:2001}.

To compare with experiment, the rate (\ref{rate}) has to be convoluted with a fireball expansion. Assuming thermalisation, we use a setup reminiscent of hydrodynamics, constraining the final state by hadronic measurements and the initial state by geometry. The evolution inbetween is taken to proceed isentropically and fixed by the EOS of the quasiparticle model. We find an initial temperature of 300 MeV and a life time of the QGP phase of 7 fm/$c$ for SPS 30\% central Pb(158 AGeV)+Au collisions. In the left panel of figure \ref{figure4}, we show the agreement of the final result with the experimental data from the CERES/NA45 collaboration. Data at a lower beam energy of 40 AGeV and $p_T$-separated data are also well described \cite{Renk:2001}.

\section{Charm suppression}
%
%
\begin{figure}[tbh]
%
\begin{center}
\vspace{-1.3cm}
\epsfig{file =  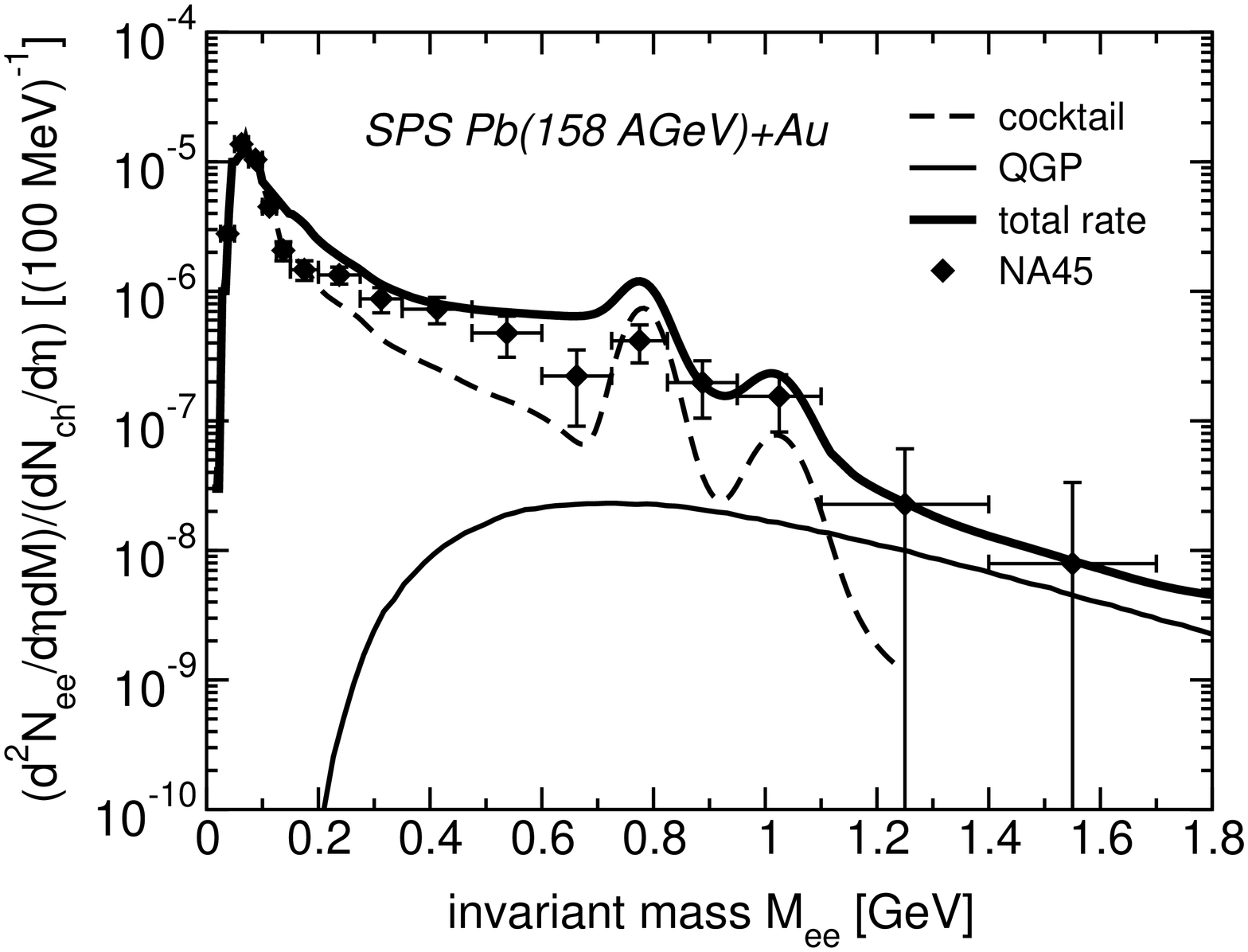,width=6.7cm, angle = -90}\raisebox{-7mm}{\epsfig{file=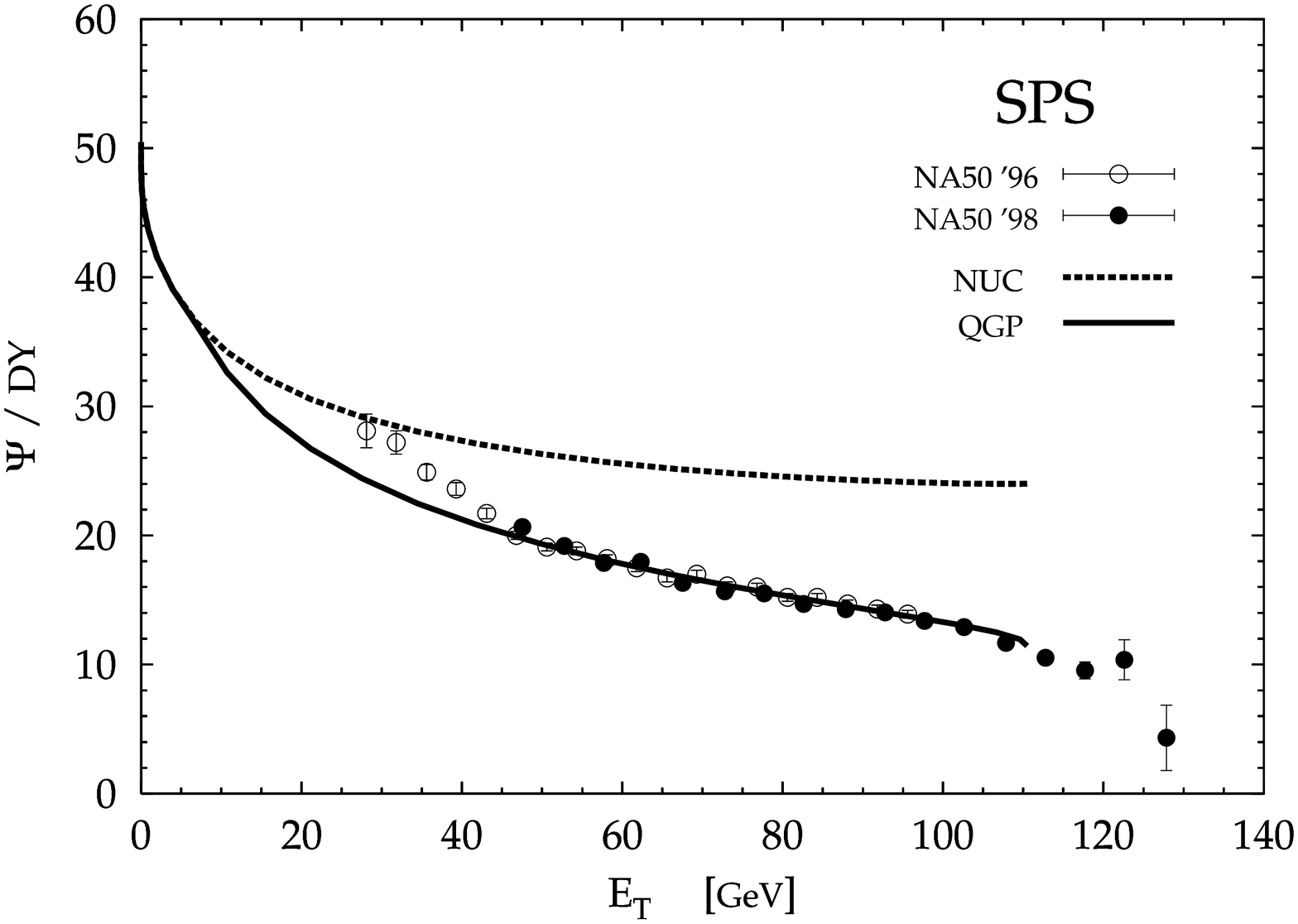,width=5.7cm, angle = -90}}
\vspace{-0.9cm}
\caption{Left panel: Dilepton rate as a function of invariant mass $M_{ee}$. Right panel: $J / \psi$ suppression, normalised to the Drell-Yan yield, as a function of transverse energy $E_T$.} \label{figure4}
\end{center}
\vspace{-0.9cm}
\end{figure}
Next, we analyse  $J / \psi$ suppression within kinetic theory \cite{Polleri:2002}, assuming that $J / \psi$s are formed in the initial state of the collision and subsequently  propagate through the thermalised medium. Interactions with the medium constituents lead to the break-up of the bound $c \bar c$ state. Neglecting coalescence, the time evolution of the $J / \psi$ number $N^y_{J/ \psi}$ at midrapidity is then described by
\begin{equation}
\frac{d}{d\tau} N^y_{J/\psi} = - \sum_n\ \langle\langle \sigma^n_D\,v_{\rm rel}
\rangle\rangle (\tau) \ \rho_n(\tau) \,N^y_{J/ \psi},
\end{equation}
where $\rho_n(\tau)$ is the density of the surrounding particles, and $\langle\langle \sigma^n_D\,v_{\rm rel}
\rangle\rangle (\tau)$ an averaged cross section. Since the gluon density of the quasiparticle model is an order of magnitude larger than any hadronic densities, the dominant dissociation process is $J / \psi + g \rightarrow c + \bar c$, with no need to take into account hadronic comovers. Folding now the time-dependent rate with the SPS fireball evolution (that is taken over unchanged from the dilepton discussion), we arrive at the right panel of figure \ref{figure4}. Here, our result is compared with data from the NA50 experiment. Again, the agreement with data is good in the region where thermalisation constitutes a valid concept. 
\section{Summary}
We have constructed a quasiparticle model of the QGP near $T_c$ with a phenomenological inclusion of confinement, which works well for both gluons and quarks. Lattice data at small chemical potential could be naturally interpreted by the quasiparticle structures of our model. We then used the resulting realistic equation of state to construct a fireball and described both dilepton radiation and charm suppression within a unified framework -- a first step towards consistent heavy-ion phenomenology.

%


\end{document}